\documentclass[]{article}

\usepackage{color}
\usepackage[usenames,dvipsnames,svgnames,table]{xcolor}
\usepackage{graphicx}

\small

\begin{document}

\twocolumn

\begin{center}
\fboxrule0.02cm
\fboxsep0.4cm
\fcolorbox{Brown}{Ivory}{\rule[-0.9cm]{0.0cm}{1.8cm}{\parbox{7.8cm}
{ \begin{center}
{\Large\em Perspective}

\vspace{0.5cm}

{\Large\bf The rotation of brown dwarfs}

\vspace{0.2cm}

{\large\em Aleks Scholz}


\vspace{0.5cm}

\centering
\includegraphics[width=0.30\textwidth]{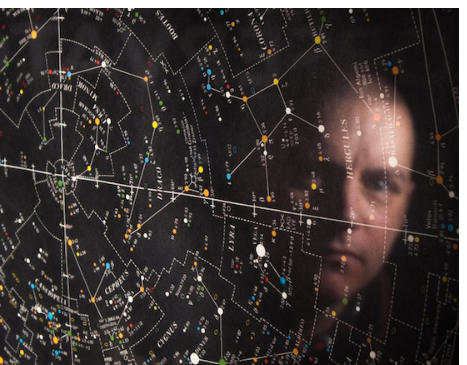}
\end{center}
}}}
\end{center}

\normalsize

{\bf 1. Introduction}

Throughout their evolution, low-mass stars lose angular momentum. A typical young star contains several orders of magnitude less angular momentum per mass unit than the molecular cloud core from which it has formed (Bodenheimer 1995). Another couple of orders of magnitude are lost on the main sequence. Through the RGB phase, the angular momentum disposal continues. The white dwarf at the end of the timeline incorporates even less specific angular momentum than the late main sequence star (although we do not quite know by how much, Kawaler 2004). Basically, angular momentum conservation is very rare in the evolution of low-mass stars.

The main agent in the rotational braking of a low-mass star is the
stellar magnetic field. During the T Tauri phase, the magnetic
interaction between star and disk effectively prevents the spin-up of
the star that would otherwise be expected due to contraction and mass infall.
For the first few million years of their existence, a significant
fraction of low-mass stars rotate with approximately constant period
(e.g., Rebull et al. 2004, Herbst \& Mundt 2005, Davies et al. 2014).
The exact physical mechanism that allows them to do that is not quite
clear, but is likely to be a combination of magnetic locking of star
and disk as well as powerful winds (see Matt et al. 2010, 2012).
Magnetically driven stellar winds control the rotation for the
remainder of the pre-main sequence and the entire main sequence phase.
On the main sequence, stellar angular momentum as well as indicators
related to magnetic activity drop approximately with the squareroot of
the age, a relation known for more than 40 years (Skumanich 1972).

Stellar rotation is interesting because it is deeply linked to many
important physical aspects of stellar evolution -- magnetic field
generation, interior structure, disk evolution, winds, internal
mixing, chemical evolution, as well as magnetic activity which has
implications for habitability of planets. Also, the rotation period,
measured from the periodic photometric modulation due to surface
features, is one of the very few fundamental stellar parameters that
can be derived for large numbers of stars with percent or better
accuracy, completely independent of models. This makes it very
tempting to use rotation as a secondary indicator to infer other
parameters, for example, to estimate stellar ages, an area of research
that has been coined gyrochronology (Barnes 2003).

Because of the propensity of low-mass stars to shed angular momentum, it is always interesting to study objects that buck the trend and manage to conserve or almost conserve angular momentum for significant periods of time. These objects give us information about the mass dependencies of the fundamental processes mentioned above, specifically, they tell us in which parameter regimes these processes fail. Planets are in this category, as are brown dwarfs, objects unable to sustain stable Hydrogen fusion and to attain thermodynamical equilibrium, with masses below 0.08$\,M_{\odot}$.  In this article I will give a summary of this exciting topic within the venerable research field of stellar rotation. I will mostly discuss the rotational evolution of {\it young} brown dwarfs (since this is the Star Formation Newsletter) and only briefly touch on the (equally exciting) topic of the rotation of evolved field brown dwarfs and the link to cloudy atmospheres (see Radigan et al. 2014, Metchev et al. 2015 and references therein)

For more details on everything that is cursorily summarised here, I refer the reader to the reviews in Protostars \& Planets V by Herbst et al. (2007) and in Protostars \& Planets VI by Bouvier et al. (2013). The former is heavily focused on observations and limited to the pre-main sequence stage, while the latter also includes a thorough overview of the theory side as well as the main-sequence evolution. 

{\bf 2. Observational timeline}

The exploration of the rotation of very low mass stars and brown dwarfs lagged only a few years behind the discovery of these objects. In the mid and late 90s, several groups reported solid evidence that objects at the bottom of the main sequence and beyond are fast rotators (e.g., Basri \& Marcy 1995, Martin 1998, Terndrup et al. 1999, Bailer-Jones \& Mundt 1999). The first {\it slowly} rotating brown dwarfs with periods of several days were published by Joergens et al. (2003) and Scholz \& Eisl{\"o}ffel (2004, 2005) for objects in young clusters and star forming regions.

\begin{figure}[t]
\centering
\includegraphics[width=0.45\textwidth]{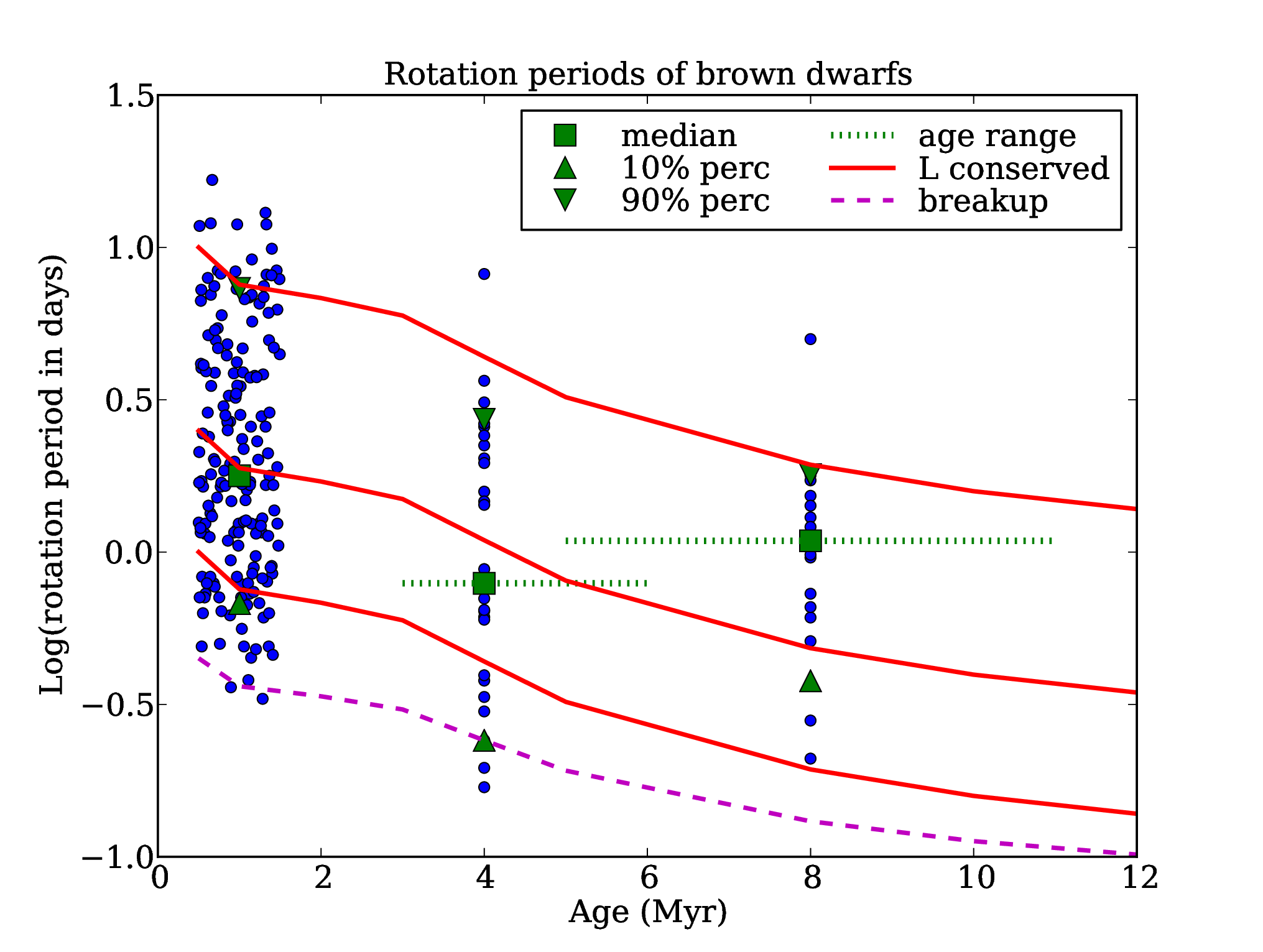}
\caption{Rotation periods for young brown dwarfs as a function of age, compared with 
evolutionary tracks assuming angular momentum conservation (solid red lines). The breakup period is 
plotted as well (dashed pink line). The green squares mark the median period for each sample; the 
green triangles the 10\% and 90\% percentile.
The periods at 1\,Myr (ONC) are randomly spread out in age for clarity.
Data from Scholz \& Eisl{\"o}ffel (2004, 2005), Rodriguez-Ledesma et al. (2009), Cody \& Hillenbrand (2010), Scholz
et al. 2015; for more details see Scholz et al. (2015).}
\label{fig1}
\end{figure}

Rotation periods in young clusters are the low hanging fruit in this
field. The dense populations of these regions allow for very
efficient, multiplexed, deep, high-cadence monitoring, using
wide-field imagers at medium-sized telescopes. In addition, young
brown dwarfs are late M dwarfs which exhibit significant magnetic spot
activity, facilitating the measurement of rotation periods.
Today we know periods for dozens substellar objects with ages
of 1-5 Myr (see Fig. \ref{fig1}), primarily thanks to a number of PhD
projects devoted to studying the variability of young stars and brown
dwarfs. This includes my own PhD work published in Scholz \&
Eisl{\"o}ffel (2004a, b, 2005, 2009, 2011), but also the work by
Markus Lamm (Lamm et al. 2004, 2005), Maria Victoria Rodriguez-Ledesma
(Rodriguez Ledesma et al. 2009, 2010) as well as Ann Marie Cody (Cody
\& Hillenbrand 2010, 2011, 2014). Some further periods for young brown
dwarfs have been published by Bailer-Jones \& Mundt (2001), Zapatero
Osorio et al. (2004), Caballero et al. (2004), and Scholz et al.
(2012). 

Earlier this year, we have used high-precision lightcurves from the revamped Kepler mission K2 to measure rotation periods for 16 young objects with masses close to or below the Hydrogen burning limit (Scholz et al. 2015, see Fig. \ref{fig1}). These are members of the Upper Scorpius star forming regions, with ages between 5 and 10 Myr, a previously unexplored age regime for brown dwarf rotation. Recently Biller et al. (2015) constrained for the first time the rotation period of a young brown dwarf with a mass below the Deuterium burning limit (lower limit of 5\,h). In all young regions studied so far, the brown dwarf periods range from a few hours up to several days. Note that some of the published periods at ages of 1-5\,Myr
are in the range of the breakup limit. If confirmed, this would have
severe implications for the further evolution of these brown dwarfs
(see Scholz \& Eisl{\"o}ffel 2005).

Measuring periods for field brown dwarfs is more difficult, because they have to be monitored one by one. Also, the origin of periodic variability is different in the older, and therefore much cooler, field brown dwarfs. Their variability is most likely caused by inhomogenuous cloud coverage ('weather'), instead of magnetic spots. While a few individual periods have been published earlier, the discovery of highly variable brown dwarfs at the L/T transition by Artigau et al. (2009) and Radigan et al. (2012) increased the interest in brown dwarf variability and motivated a number of monitoring projects for field brown dwarfs aimed at studying cloud properties, which resulted, almost as a side-product, in many known rotation periods (e.g., Radigan et al. 2014, Metchev et al. 2015). So far, the overwhelming majority of the published periods for field brown dwarfs are {\it short} -- less than 20 hours -- but there are some exceptions (see Metchev et al. 2015). Since the monitoring runs for most objects have also been short, the upper limit for brown dwarf periods remains poorly defined.

This overview is focused on photometrically measured rotation periods. In comparison with spectroscopic rotational velocities, they have the advantage of being much more accurate and free of the inclination factor. However, incompleteness and bias in period samples is always an issue, as periods can only be inferred when asymmetrically distributed surface features are present. Therefore it is important to note that the record of rotational velocities available for brown dwarfs at various ages (e.g., Bailer-Jones 2004, Zapaterio Osorio 2006, Reiners \& Basri 2008, Konopacky et al. 2012) so far  corroborates the findings from rotation periods. 

{\bf 3. Inefficient disk braking}

Disk braking is a term I like to use for whatever disk-related
mechanism that manages to keep the rotation periods of young stars 'locked'
for ages of 1 to 5 Myr. Young brown dwarfs have much lower
luminosities, lower accretion rates, lower disk masses, and (perhaps)
different magnetic field properties than typical T Tauri stars.
Testing disk braking in brown dwarfs is therefore a good way to
illuminate how the braking mechanism works in a very different
parameter regime.

The observational study of disk braking in brown dwarfs is still
crippled by the lack of decently sized samples with measured rotation
periods {\it and} mid-infrared photometry. The latter is essential to
reliably test for the presence of the disk. In Scholz \& Eisl{\"o}ffel
(2004) we used the variability amplitude as indicator for accretion
and, hence, disk, and find that the slow rotators all are accretors,
which would be evidence for disk braking, but in a small sample which
also included very low mass stars. Rodriguez Ledesma et al. (2010)
find no connection between slow rotation and presence of the disk in
their sample of ONC brown dwarfs, but they use near-infrared
photometry as disk indicators. At near-infrared wavelengths, the
photospheric emission of brown dwarf peaks and the disk is faint,
which makes it very difficult to robustly detect the disk. Cody \&
Hillenbrand (2010) test the disk-rotation relation in the substellar
regime for the first time with mid-infrared photometry and do not see
evidence for a direct connection between rotation rate and the
presence of a disk -- but their sample only contains few brown dwarfs.

In our most recent study, we use the periods derived for brown dwarfs
in Upper Scorpius in combination with literature data to trace for the
first time the substellar rotational evolution from 1 to 10\,Myr
(Scholz et al. 2015). The 'money plot' from this paper is reproduced
in Fig. \ref{fig1}, which simply shows periods vs. age in comparison
with tracks assuming angular momentum conservation. We find that the
period evolution over this age span is consistent with no angular
momentum loss, i.e. no rotational braking at all. If disk braking
occurs, the locking timescale is at most 2-3 Myr, significantly
shorter than for low-mass stars. This finding is robust within the
uncertainties for the cluster ages. However, we also identify the
disk-bearing objects in our sample, using WISE photometry, and find
that all objects in our sample which still harbour disks are among the
slowest rotators. Thus, while disk braking seems to be very
inefficient in brown dwarfs, as already found by Cody \& Hillenbrand
(2010) (and by Lamm et al. 2005 for slightly more massive objects),
it does seem to be at work at least in a few selected objects which
manage to retain their disks longest.

Since disk lifetimes in brown dwarfs are not vastly different from stars (see Dawson et al. 2013), these findings indicate that the interaction between star and disk changes as we go to very low masses. Low ionisation at the inner disk edge (due to lower luminosity), lower mass accretion rates, as well as changes in the magnetic configuration all seem plausible explanations at this stage, but this needs to be explored in detail. In any case, whatever braking mechanism is at work in low-mass T Tauri stars, it should become inefficient in the substellar regime.

{\bf 4. Also: inefficient wind braking}

Wind braking controls the long-term rotational evolution of solar-mass
stars. By the time stars with spectral type F to K have reached
the age of the Hyades (600 Myr), they have settled onto a well defined
period-mass relationship, a kind of 'main sequence' of rotational
evolution. At this point the rotation period of low-mass stars is a
function of mass and age, and little else (at least once binary stars
in tidal interaction have been eliminated). With decreasing mass, the
time objects need to converge to the rotational main sequence
increases substantially (Irwin et al. 2011, Scholz et al. 2012, Newton
et al. 2015); at 0.1-0.3$\,M_{\odot}$, the spindown timescale is in
the range of gigayears. The origin of this mass dependence is not well
understood, but is likely related to the details of the wind physics
and/or magnetic field generation. 

The fast rotation of most field brown dwarfs indicates that this trend continues in the brown dwarf regime. Extrapolating from the evolutionary tracks for very low mass stars, we would expect brown dwarfs to retain their fast rotation rates for more than 5\,Gyr. In addition to changes in magnetic field properties, most of the atmospheres of brown dwarfs eventually become too cool for an efficient coupling between plasma and magnetic field (Mohanty et al. 2003, Rodriguez-Barrera et al. 2015), which mostly shuts down persistent H$\alpha$ and X-ray activity (apart from transient events) and may further impede rotational braking. Based on the currently known periods, however, there is evidence for {\it some} rotational braking on long timescales, but the angular momentum loss rate may be $\sim 10000$ times weaker than in solar-mass stars (Bouvier et al. 2013). Large and unbiased period samples of field brown dwarfs at various ages are needed to constrain this further. At this stage, however, it seems clear that wind braking in brown dwarfs does not work very well.

{\bf 5. A universal spin-mass relation?}

\begin{figure}[t]
\centering
\includegraphics[width=0.45\textwidth]{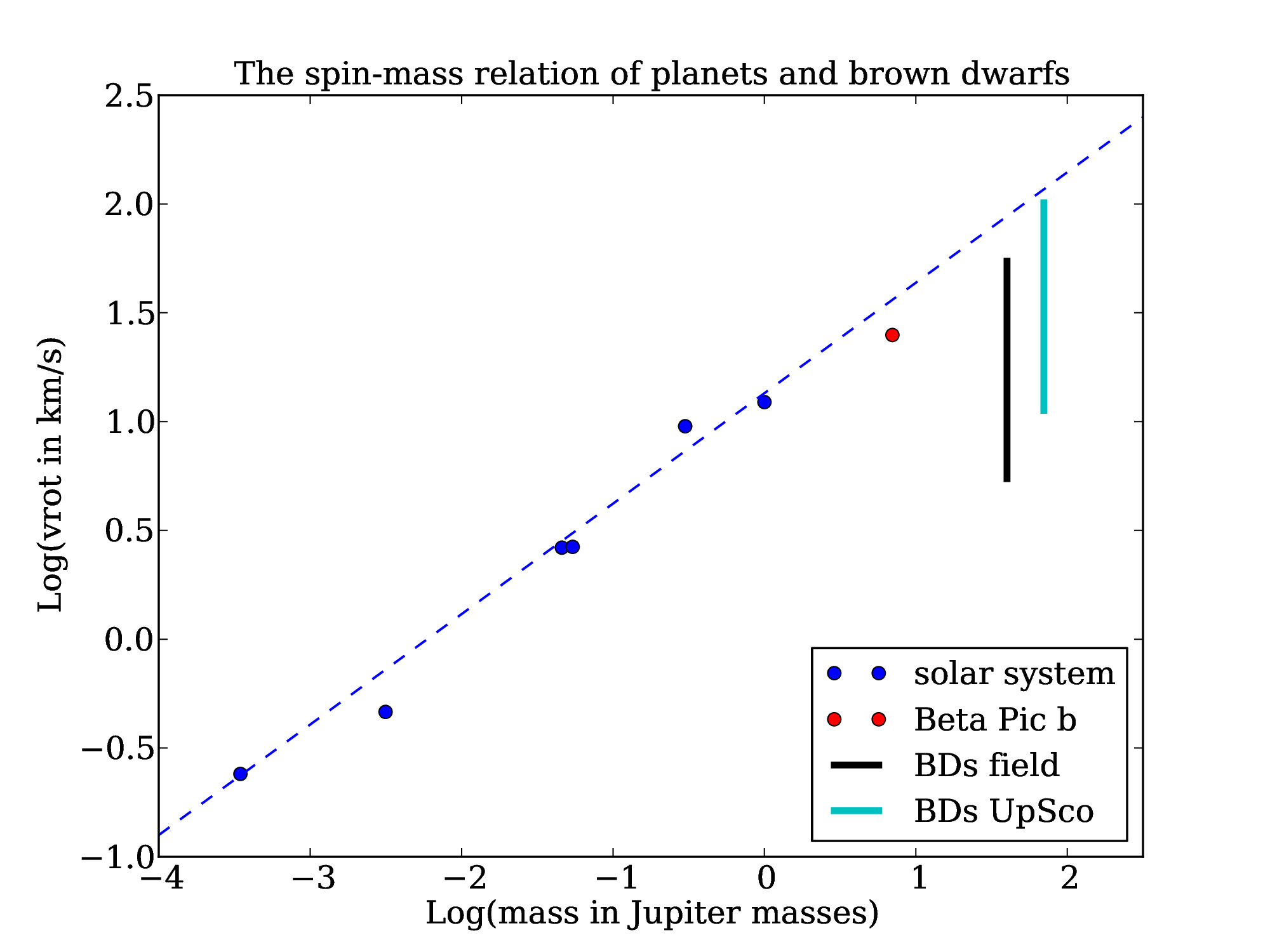}
\caption{Equatorial rotational velocity (in km/s) vs. mass (in Jupiter masses) for 
brown dwarfs in comparison with solar system planets (excluding Venus and Mercury) 
and $\beta$\,Pic\,b. For brown dwarfs we show the typical range of rotational velocities.
Unpublished figure, adapted from Snellen et al. (2014). The dashed line is not a fit,
but only overplotted to illustrate the planetary spin-mass relation.}
\label{fig2}
\end{figure}

In terms of their rotational evolution, brown dwarfs are much more like giant planets than stars -- they retain their initial rotation rates for cosmological timescales. Therefore, it makes sense to compare rotation rates of brown dwarfs with those of planets. Solar system planets which are not tidally spun down, e.g. Mars, Jupiter, Saturn, Uranus, Neptune, obey a surprisingly strict relation between equatorial rotational velocity and mass (e.g., Hughes 2003), shown in Fig. \ref{fig2}. The Earth falls slightly below this correlation due to tidal interaction with the Moon, but fits the line fairly well when the Moon 'is put back into Earth' (Hughes 2003). This impressive relation can also be extended towards lower masses by including asteroidal data. As shown by Snellen et al. (2014), the rotational velocity for the young, massive exoplanet $\beta$\,Pic\,b also falls on this spin-mass relation, at least given the errors on its mass. Note that a similar trend is also apparent when plotting angular momentum or specific angular momentum instead of rotational velocity. The relation is likely to be directly linked to the formation processes, presumably to the growth of planets in the protoplanetary disk.

In Fig. \ref{fig2}, I overplot the typical range of rotational velocities, inferred from periods, for field brown dwarfs (black line, based on a period range from 2 to 20\,h, see Metchev et al. 2015) and for young Upper Scorpius brown dwarfs (magenta line, based on a period range 0.2 to 1.8\,d, see Scholz et al. 2015). For this plot, I used fiducial masses of 40 and 70 Jupiter masses, but the result does not depend on that choice. It seems clear that the brown dwarf populations, young and old, violate the planetary spin-mass relation. Only the fastest rotating young brown dwarfs, with periods of 5-10\,h, and maybe some ultrafast rotating field dwarfs (Konopacky et al. 2012), might be close to the spin-mass relation. The remaining objects are clearly below the trend observed for planets, by up to one order of magnitude.

This could be an effect of formation or evolution or both. The consensus view is that most brown dwarfs form in the way similar to stars, as a by-product of cloud collapse and fragmentation (see Luhman 2014, Scholz et al. 2012). In contrast to planets, the specific angular momentum of low-mass stars is not a strong function of mass (also an interesting constraint on formation), and brown dwarfs seem to fit into that trend (Herbst et al. 2001, Cody \& Hillenbrand 2010, Wolff et al. 2004). Therefore it is maybe not surprising that most brown dwarfs violate the planetary spin-mass relation, even at young ages. It is then conceivable that the few brown dwarfs that {\it do} obey the spin-mass relation might be the exceptions that form like giant planets and are ejected at an early evolutionary stage from the disk. If this relation really holds for all young substellar objects formed in disks, a hypothesis we need to test further, this could help us to distinguish brown dwarf formation scenarios.

On the other hand, brown dwarfs do spin down, albeit slowly, which makes the interpretation of Fig. \ref{fig2} more difficult. Planets not affected by tidal interactions might be the only objects in the Galaxy that retain their primordial rotation rate and conserve angular momentum. By including brown dwarfs in the spin-mass diagram, we see the subtle emergence of the mechanisms that are responsible for the efficient spindown of stars.

\footnotesize

{\bf References:}

Artigau, E., et al., 2009, ApJ, 701, 1534\\
Bailer-Jones, C. A. L., Mundt, R., 1999, A\&A, 348, 800\\
Bailer-Jones, C. A. L., Mundt, R., 2001, A\&A, 367, 218\\
Bailer-Jones, C. A. L., 2004, A\&A, 419, 703\\
Barnes, S., 2003, ApJ, 586, 464\\
Basri, G., Marcy, G. W, 1995, AJ, 109, 762\\
Biller, B. E., et al., 2015, ApJ, 813, 23\\
Bodenheimer, P., 1995, ARA\&A, 33, 199\\
Bouvier, J., et al., 2014, Protostars \& Planets VI, 433\\
Caballero, J. A., et al., 2004, A\&A, 424, 857\\
Cody, A. M., Hillenbrand, L. A., 2010, ApJS, 191, 389\\
Cody, A. M., Hillenbrand, L. A., 2011, ApJ, 741, 9\\
Cody, A. M., Hillenbrand, L. A., 2014, ApJ, 796, 129\\
Davies, C. L., et al., 2014, MNRAS, 444, 1157\\
Dawson, P., et al., 2013, MNRAS, 429, 903\\
Herbst, W., Mundt, R., 2005, ApJ, 633, 967\\
Herbst, W., et al., 2007, Protostars \& Planets V, 297\\
Hughes, D. W., 2003, P\&SS, 51, 517\\
Irwin, J., et al., 2011, ApJ, 727, 56\\
Joergens, V., et al., 2003, ApJ, 594, 971\\
Kawaler, S. D., 2004, IAUS, 215, 561\\
Konopacky, Q. M., 2012, ApJ, 750, 79\\
Lamm, M. H., et al., 2004, A\&A, 417, 557\\
Lamm, M. H., et al., 2005, A\&A, 430, 1005\\
Luhman, K. L., 2012, ARA\&A, 50, 65\\
Martin, E. L., Zapatero Osorio, M. R., 1997, MNRAS, 286, 17\\
Matt, S. P., et al., 2010, ApJ, 714, 989\\
Matt, S. P., et al., 2012, ApJ, 745, 101\\
Metchev, S. A., et al., 2015, ApJ, 799, 154\\
Mohanty, S., Basri, G., 2003, ApJ, 583, 451\\
Newton, E. R., et al., 2015, arXiv, 51100957\\
Radigan, J., et al., 2012, ApJ, 750, 105\\
Radigan, J., et al., 2014, ApJ, 793, 75\\
Rebull, L. M., et al., 2004, AJ, 127, 1029\\
Reiners, A., Basri, G., 2010, ApJ, 710, 924\\
Reiners, A., Basri, G., 2008, ApJ, 684, 1390\\
Rodriguez-Barrera, M. I., et al., 2015, MNRAS, 454, 3977\\
Rodriguez-Ledesma, M. V., et al., 2009, A\&A, 502, 883\\
Rodriguez-Ledesma, M. V., et al., 2010, A\&A, 515, 13\\
Scholz, A., Eisl{\"o}ffel, J., 2004, A\&A, 419, 249\\
Scholz, A., Eisl{\"o}ffel, J., 2004, A\&A, 421. 259\\
Scholz, A., Eisl{\"o}ffel, J., 2005, A\&A, 427, 1007\\
Scholz, A., Eisl{\"o}ffel, J., 2007, MNRAS, 381, 1638\\
Scholz, A., et al., 2009, MNRAS, 400, 1548\\
Scholz, A., et al., 2011, MNRAS, 413, 2595\\
Scholz, A., et al., 2012, MNRAS, 419, 1271\\
Scholz, A., et al., 2012, ApJ, 756, 24\\
Scholz, A., et al., 2015, ApJ, 809, 29\\
Skumanich, A., 1972, ApJ, 171, 565\\
Snellen, I., et al., 2014, Nature, 509, 63\\
Terndrup, D. M., 1999, AJ, 118, 1814\\
Wolff, S. C., 2004, ApJ, 601, 979\\ 
Zapatero Osorio, M. R., et al., 2003, A\&A, 408, 663\\
Zapatero Osorio, M. R., et al., 2006, ApJ, 647, 1405


\normalsize

\end{document}